\documentclass[onecolumn,11pt]{article}
\usepackage[top=1in, bottom=1in, left=1in, right=1in]{geometry}
\usepackage{amsmath,amsfonts,amscd,amssymb}
\usepackage{graphicx}
\graphicspath{{figures/}}
\usepackage{url}
\usepackage{xcolor}
\usepackage{caption}
\usepackage{subcaption}
\usepackage{mathtools}
\mathtoolsset{showonlyrefs}
\usepackage{setspace}
\usepackage{tikz}
\usetikzlibrary{shapes,arrows,calc}
\usepackage[numbers,sort&compress]{natbib}

\usepackage{listings}
\usepackage{xcolor}

\definecolor{maroon}{cmyk}{0, 0.87, 0.68, 0.32}
\definecolor{halfgray}{gray}{0.55}
\definecolor{ipython_frame}{RGB}{207, 207, 207}
\definecolor{ipython_bg}{RGB}{247, 247, 247}
\definecolor{ipython_red}{RGB}{186, 33, 33}
\definecolor{ipython_green}{RGB}{0, 128, 0}
\definecolor{ipython_cyan}{RGB}{64, 128, 128}
\definecolor{ipython_purple}{RGB}{170, 34, 255}

\lstdefinelanguage{iPython}{
    morekeywords={access,and,break,class,continue,def,del,elif,else,except,exec,finally,for,from,global,if,import,in,is,lambda,not,or,pass,print,raise,return,try,while,True,False,as},%
    morekeywords=[2]{abs,all,any,basestring,bin,bool,bytearray,callable,chr,classmethod,cmp,compile,complex,delattr,dict,dir,divmod,enumerate,eval,execfile,file,filter,float,format,frozenset,getattr,globals,hasattr,hash,help,hex,id,input,int,isinstance,issubclass,iter,len,list,locals,long,map,max,memoryview,min,next,object,oct,open,ord,pow,property,range,raw_input,reduce,reload,repr,reversed,round,set,setattr,slice,sorted,staticmethod,str,sum,super,tuple,type,unichr,unicode,vars,xrange,zip,apply,buffer,coerce,intern},%
    sensitive=true,%
    morecomment=[l]\#,%
    morestring=[b]',%
    morestring=[b]",%
    morestring=[s]{'''}{'''},
    morestring=[s]{"""}{"""},
    morestring=[s]{r'}{'},
    morestring=[s]{r"}{"},%
    morestring=[s]{r'''}{'''},%
    morestring=[s]{r"""}{"""},%
    morestring=[s]{u'}{'},
    morestring=[s]{u"}{"},%
    morestring=[s]{u'''}{'''},%
    morestring=[s]{u"""}{"""},%
    literate=
    {á}{{\'a}}1 {é}{{\'e}}1 {í}{{\'i}}1 {ó}{{\'o}}1 {ú}{{\'u}}1
    {Á}{{\'A}}1 {É}{{\'E}}1 {Í}{{\'I}}1 {Ó}{{\'O}}1 {Ú}{{\'U}}1
    {à}{{\`a}}1 {è}{{\`e}}1 {ì}{{\`i}}1 {ò}{{\`o}}1 {ù}{{\`u}}1
    {À}{{\`A}}1 {È}{{\'E}}1 {Ì}{{\`I}}1 {Ò}{{\`O}}1 {Ù}{{\`U}}1
    {ä}{{\"a}}1 {ë}{{\"e}}1 {ï}{{\"i}}1 {ö}{{\"o}}1 {ü}{{\"u}}1
    {Ä}{{\"A}}1 {Ë}{{\"E}}1 {Ï}{{\"I}}1 {Ö}{{\"O}}1 {Ü}{{\"U}}1
    {â}{{\^a}}1 {ê}{{\^e}}1 {î}{{\^i}}1 {ô}{{\^o}}1 {û}{{\^u}}1
    {Â}{{\^A}}1 {Ê}{{\^E}}1 {Î}{{\^I}}1 {Ô}{{\^O}}1 {Û}{{\^U}}1
    {œ}{{\oe}}1 {Œ}{{\OE}}1 {æ}{{\ae}}1 {Æ}{{\AE}}1 {ß}{{\ss}}1
    {ç}{{\c c}}1 {Ç}{{\c C}}1 {ø}{{\o}}1 {å}{{\r a}}1 {Å}{{\r A}}1
    {€}{{\EUR}}1 {£}{{\pounds}}1,
    literate=
    *{-}{{{\color{ipython_purple}-}}}1
     {?}{{{\color{ipython_purple}?}}}1
     {^}{{{\color{ipython_purple}\^{}}}}1
     {=}{{{\color{ipython_purple}=}}}1
     {+}{{{\color{ipython_purple}+}}}1
     {*}{{{\color{ipython_purple}$^\ast$}}}1
     {/}{{{\color{ipython_purple}/}}}1
     {>}{{{\color{ipython_purple}$>$}}}1
     {<}{{{\color{ipython_purple}$<$}}}1
     {@}{{{\color{ipython_purple}@}}}1
     {+=}{{{+=}}}1
     {-=}{{{-=}}}1
     {*=}{{{$^\ast$=}}}1
     {/=}{{{/=}}}1,
    identifierstyle=\color{black}\ttfamily,
    commentstyle=\color{ipython_cyan}\ttfamily,
    stringstyle=\color{ipython_red}\ttfamily,
    keepspaces=true,
    showspaces=false,
    showstringspaces=false,
    rulecolor=\color{ipython_frame},
    backgroundcolor=\color{ipython_bg},
    basicstyle=\normalsize,
    keywordstyle=\color{ipython_green}\ttfamily,
}

\definecolor{mygray}{gray}{0.95}
\definecolor{codegreen}{rgb}{0,0.6,0}
\definecolor{codegray}{rgb}{0.5,0.5,0.5}
\definecolor{codepurple}{rgb}{0.58,0,0.82}
\definecolor{backcolour}{rgb}{0.95,0.95,0.92}

\lstdefinestyle{mystyle}{
    backgroundcolor=\color{mygray},   
    commentstyle=\color{codegreen},
    keywordstyle=\color{magenta},
    numberstyle=\tiny\color{codegray},
    stringstyle=\color{codepurple},
    basicstyle=\ttfamily\footnotesize,
    breakatwhitespace=false,         
    breaklines=true,                 
    captionpos=b,                    
    keepspaces=true,                 
    showspaces=false,                
    showstringspaces=false,
    showtabs=false,                  
    tabsize=2
}

\usepackage[bottom,flushmargin,hang,multiple]{footmisc}
\usepackage{lipsum}
\newcommand\blfootnote[1]{%
  \begingroup
  \renewcommand\thefootnote{}\footnote{#1}%
  \addtocounter{footnote}{-1}%
  \endgroup
}

\definecolor{header1}{cmyk}{0,0,0,1}

\DeclareMathOperator*{\argmin}{arg\rm{}min}

\newcommand{\Cv}{\mathbf{C}}

\newcommand{\xv}{\mathbf{x}}
\newcommand{\yv}{\mathbf{y}}

\newcommand{\bbm}{\begin{bmatrix}}
\newcommand{\ebm}{\end{bmatrix}}
\newcommand{\horzline}{\rule[.5ex]{1.5em}{0.4pt}}

\tikzstyle{block} = [rectangle, draw, fill=blue!20, 
    text width=4.5em, text centered, rounded corners, minimum height=4em]
\tikzstyle{line} = [draw, -latex']

\title{\vspace{-.65in}{\fontsize{16}{16}\selectfont \textbf{PySensors: A Python Package for Sparse Sensor Placement}}\vspace{-.15in}}

\author{\normalsize{
Brian M. de Silva$^{1*}$, Krithika Manohar$^2$, Emily Clark$^{3}$,}\\ \normalsize{Bingni W. Brunton$^4$, Steven L. Brunton$^2$, J. Nathan Kutz$^1$}\\
\footnotesize{$^1$ Department of Applied Mathematics, University of Washington, Seattle, WA 98195, United States}\\
\footnotesize{$^2$ Department of Mechanical Engineering, University of Washington, Seattle, WA 98195, United States}\\
\footnotesize{$^3$ Department of Physics, University of Washington, Seattle, WA 98195, United States}\\
\footnotesize{$^4$ Department of Biology, University of Washington, Seattle, WA 98195, United States \vspace{-.2in}}
}
\date{}
\begin{document}
\maketitle

\blfootnote{$^*$ Corresponding author (bdesilva@uw.edu).}
\vspace{-.2in}
\begin{abstract}
\texttt{PySensors} is a Python package for selecting and placing a sparse set of sensors for classification and reconstruction tasks.  Specifically, 
\texttt{PySensors} implements algorithms for data-driven {\em sparse sensor placement optimization for reconstruction} (SSPOR)~\cite{manohar2018data} and {\em sparse sensor placement optimization for classification} (SSPOC)~\cite{brunton2016sparse}. In this work we provide a brief description of the mathematical algorithms and theory for sparse sensor optimization, along with an overview and demonstration of the features implemented in \texttt{PySensors} (with code examples).
We also include practical advice for user and a list of potential extensions to \texttt{PySensors}.  
Software is available at \url{https://github.com/dynamicslab/pysensors}.

\vspace{0.05in}
\vspace{-.15in}
\end{abstract}

\section{Introduction}

The success of predictive models and controllers in engineering and natural processes is largely determined by critical {\em in situ} measurements and feedback from sensors~\cite{Brunton2019book}.  However, the deployment of sensors into complex environments, such as manufacturing~\cite{Manohar2018jms}, geophysical~\cite{Yildirim:2009} and biological processes~\cite{colvert2017local,Mohren2018pnas}, is often expensive and challenging. Moreover, performance outcomes are extremely sensitive to the location and number of sensors deployed, motivating the optimal placement of sensors for diverse decision-making tasks.  In general, choosing globally optimal placements within the search space of a large-scale complex system is an intractable computation, in which the number of possible placements grows combinatorially with the number of candidates~\cite{ko1995exact}. While sensor placement, or sensor selection, has traditionally been guided by expert knowledge and first principles models, the continued growth in system complexity has motivated new mathematical paradigms and optimization algorithms for data collection and data-driven modeling that aim to automate the optimal, or near-optimal, sensor selection task.

A number of automated sensor placement methods have been developed in recent years, designed to optimize outcomes in the design of experiments~\cite{Boyd2004convexbook,joshi2008sensor}, convex~\cite{joshi2008sensor,brunton2016sparse} and submodular objective functions~\cite{summers2015submodularity}, information theoretic and Bayesian criteria~\cite{Caselton1984spl,krause2008near,Lindley1956ams,Sebastiani2000jrss,Paninski2005nc}, optimal control~\cite{Dhingra2014cdc,Munz2014ieeetac,Zare2018arxiv,Manohar2018arxivB}, for sampling and estimating signals over graphs~\cite{Ribeiro2010sigcomm,DiLorenzo2016ieee,Chen2016ieee,Chepuri2016sam}, and reduced order modeling~\cite{Barrault2004crm,willcox2006unsteady,Chaturantabut2010siamjsc,Chaturantabut2012siamjna,drmac2016siam,manohar2018data,clark2018greedy}.  For the most part, these algorithms circumvent the computationally intractable combinatorial optimization procedure required for the globally optimal placement of sensors by positing greedy algorithms which are near-optimal and computationally efficient.  Thus near-optimal performance can be achieved with fast algorithms.

\texttt{PySensors} is a Python package for the scalable optimization of sensor placements from data. In particular, \texttt{PySensors} provides tools for sparse sensor placement optimization approaches that employ data-driven dimensionality reduction ~\cite{brunton2016sparse,manohar2018data}. This approach results in near-optimal placements for various decision-making tasks and can be readily customized using different optimization algorithms and objective functions. 
The \texttt{PySensors} package is aimed at researchers and practitioners alike, enabling anyone with access to measurement data to engage in scientific model discovery. The package is designed to be accessible to inexperienced users, adhering to \texttt{scikit-learn} standards, while also including customizable options for more advanced users.  
A number of popular sensor placement variants are implemented, but \texttt{PySensors} is also designed to enable further extensions for research and experimentation.

Maximizing the impact of sensor placement algorithms requires tools to make them accessible to scientists and engineers across various domains and at various levels of mathematical expertise and sophistication. \texttt{PySensors} unifies the algorithms developed in the recent papers~\cite{manohar2018data,clark2018greedy,brunton2016sparse} and their accompanying codes \texttt{SSPOR\_pub} and \texttt{SSPOC\_pub} into one software package. The only other packages in this domain of which we are aware are \texttt{Chama} \cite{klise2017sensor} and \texttt{Polire} \cite{narayanan2020toolkit}. While these packages and \texttt{PySensors} all enable sparse sensor placement optimization, \texttt{Chama} and \texttt{Polire} are geared towards event detection and Gaussian processes respectively, whereas \texttt{PySensors} is aimed at signal reconstruction and classification tasks.
As such, there are marked differences in the objective functions optimized by \texttt{PySensors} and its precursors.
In addition to these two packages, researchers and practitioners have made available various custom scripts for sensor placement. 
Currently, researchers seeking to employ modern sensor placement methods must choose between implementing them from scratch or manually augmenting existing unpolished codes.

\section{Background}\label{sec:background}
\texttt{PySensors} was designed to solve \textit{reconstruction} and \textit{classification} tasks, which often arise in the modeling, prediction, and control of complex processes in geophysics, fluid dynamics, biology, and manufacturing.

\subsection{Reconstruction}
\texttt{PySensors} implements the {\em sparse sensor placement optimization for reconstruction} (SSPOR)
method for recovering high-dimensional signals $\xv$ from linear sensor measurements of the form
\begin{equation}\label{eq:dynamical_system}
  \mathbf{y} = \mathbf{Cx}. 
\end{equation}
Given data in the form of state measurements $\xv_k \in \mathbb{R}^n , k=1,\dots,m$, SSPOR identifies the optimal measurements of $\xv$, given by the operator $\Cv$, which describes which components of $\xv$ to observe.  The SSPOR framework aims to find the best subset of the available measurements (usually, components of $\xv$) from which the full signal can be recovered in the estimation problem 
\begin{equation}
\Cv^\star = \argmin_{\Cv} \|\xv - f(\Cv\yv)\|_2^2
\end{equation}
We assume that $\Cv$ is a mostly sparse subset selection operator consisting of rows of the identity, with nonzero entries designating the selected measurements. Given a $p$ sensor budget and $n$ candidates state components, $\Cv$ is constrained to have the following structure
\begin{equation}
    \Cv = \begin{bmatrix} \mathbf{e}_{\gamma_1}^\top & \mathbf{e}_{\gamma_2}^\top & \dots & \mathbf{e}_{\gamma_p}^\top \end{bmatrix}, 
\end{equation}
where $\mathbf{e}_j$ is the canonical basis vector with a unit entry at the $j$th component and zeros elsewhere. The action of this measurement operator extracts the selected components of the signal 
\begin{equation}
    \yv= \Cv\xv = [x_{\gamma_1}, x_{\gamma_2}, \dots ,x_{\gamma_p}]^\top. 
\end{equation}

\subsection{Classification}

The SSPOC formulation seeks the placement of a small number of point sensors that classify high-dimensional signals $\mathbf{x} \in \mathbb{R}^n$ as one of $c$ classes. 
We start with labeled training data $\{(\mathbf{x}_i, y_i) | i=1,2,\dots,m\}$, where training examples $\mathbf{x}_i\in\mathbb{R}^n$ are concatenated into a matrix $\mathbf{X}\in\mathbb{R}^{n\times m}$ and labels $y_i\in \{0, 1, \dots, c-1\}$ are collected in a vector $\mathbf{y}\in\mathbb{R}^m$.

An $r$-dimensional feature basis of the training data $\mathbf{\Psi}\in\mathbb{R}^{n\times r}$ is extracted, where $r \ll m$.
Next, a decision space that best separates classes of training data $\mathbf{w}$ is computed on training data projected onto $\mathbf{\Psi}$. $\mathbf{w}$ is obtained as the weights of a linear classifier fit to predict labels $\mathbf{y}$ from features $\mathbf{\Psi}^\dagger \mathbf{X}$.
SSPOC seeks a measurement vector $\mathbf{s}$ that satisfies $\mathbf{\Psi}^\dagger \mathbf{s} = \mathbf{w}$, where $\mathbf{\Psi}^\dagger$ denotes the Moore-Penrose pseudoinverse of $\mathbf{\Psi}$.

In particular, we seek the sparse solution $\mathbf{s}$
\begin{equation}
    \mathbf{s} = \argmin_{\mathbf{s'}} \| \mathbf{s'} \|_1,~~~~\text{subject to}~~ \|\mathbf{\Psi}^\dagger \mathbf{s'} - \mathbf{w}\|_F < \epsilon,
\end{equation}
where $\epsilon$ is a small error tolerance.
The non-zero elements of $\mathbf{s}$ are the sensor locations. 

\section{Features}
    \texttt{PySensors} enables the sparse placement of sensors for two classes of problems: reconstruction and classification. Each problem has a dedicated class with a number of options for tailoring the package to different applications of interest.
    Additionally, \texttt{PySensors} provides methods to enable straightforward exploration of the impacts of critical hyperparameters like the number of sensors or basis modes.
    Furthermore, because \texttt{PySensors} was built with \texttt{scikit-learn} compatibility in mind, it is easy to use cross-validation to select among possible choices of bases, basis modes, and other hyperparameters.
    
    The package is divided into three primary submodules: \texttt{reconstruction}, \texttt{classification}, and \texttt{basis}. The \texttt{reconstruction} and \texttt{classification} submodules contain classes and methods for reconstruction and classification problems, respectively. And \texttt{basis} houses different implementations of various bases. There are also two helper submodules: 
    \texttt{optimizers} holds optimization routines used by the sensor selectors and \texttt{utils} has a variety of utility functions. In the subsections that follow we cover each of the main submodules in more depth.

\subsection{Reconstruction}
    For reconstruction problems the package implements a unified \texttt{SSPOR} class (SSPOR is an acronym for Sparse Sensor Placement Optimization for Reconstruction), with methods for efficiently analyzing the effects that data or sensor quantity have on reconstruction performance~\cite{manohar2018data}.
    When a \texttt{SSPOR} instance is fit to measurement data, internally it first fits a basis object to the data (see Section \ref{sec:basis}), then employs the computationally efficient and flexible QR algorithm~\cite{duersch2015true,martinsson2015blocked,Martinsson2017siamjsc}, which has recently been used for hyperreduction in reduced-order modeling~\cite{drmac2016siam} and for sparse sensor selection~\cite{manohar2018data}.
    The learned sensors are \textit{specific} to the basis that was chosen; the method leverages structure present in the basis to decide which sensors are most important.
    
    Often different sensor locations impose variable costs, e.g. if measuring sea-surface temperature, it may be more expensive to place buoys/sensors in the middle of the ocean than close to shore.
    These costs can be taken into account during sensor selection via a built-in cost-sensitive optimization routine~\cite{clark2018greedy}. This \texttt{CCQR} algorithm is found in the \texttt{optimizers} submodule along with the standard \texttt{QR} algorithm.

\subsection{Classification}
    For classification tasks, the package implements the Sparse Sensor Placement Optimization for Classification (SSPOC) algorithm~\cite{brunton2016sparse}, allowing one to optimize sensor placement for classification accuracy. 
    The algorithm is related to the compressed sensing optimization~\cite{Candes2006cpam,Donoho2006ieeetit,Baraniuk2007ieeespm}, but identifies the sparsest set of sensors that reconstructs a discriminating plane in a feature subspace.
    To instantiate a \texttt{SSPOC} object, one specifies a basis and a linear classifier.
    The implementation is fully general in the sense that it can be used in conjunction with any linear classifier and any basis from the \texttt{basis} submodule, however a linear discriminant analysis (LDA) classifier and \texttt{Identity} basis are used by default.
    When the fit method is called, a \texttt{SSPOC} instance (a) fits a basis object to the data, (b) fits the classifier to a set of examples dependent on the newly learned basis, (c) solves an optimization problem involving the weights of the classifier and the basis, (d) the sensors are selected based on the output of (c), and (e) the classifier is optionally refit using data sampled at the chosen sensor locations. The learned sensors depend on the combination of basis and classifier and on additional hyperparameters.
    
    \texttt{SSPOC} employs different optimization methods depending on whether a binary or multi-class classification problem is being solved. The \texttt{Scikit-learn} orthogonal matching pursuit implementation is used to solve binary problems and the multi-task Lasso implementation is used for multi-class problems. Note that the CVX\footnote{\url{http://cvxr.com/cvx/}} package was used in the original SSPOC formulation~\cite{brunton2016sparse}.

\subsection{Basis}\label{sec:basis}
    It is well known~\cite{manohar2018data} that the basis in which one represents measurement data can have a pronounced effect on the sensors that are selected and the quality of the reconstruction.
    Users can readily switch between different bases typically employed for sparse sensor selection:
    \begin{itemize}
        \item \texttt{Identity}: Use the raw measurement data directly, without modification. This class also empowers the user to work with other bases than those provided by \texttt{PySensors}---one can map the data to a different basis before feeding it to a \texttt{PySensors} class instantiated with an \texttt{Identity} basis.
        \item \texttt{SVD}: Use the left singular vectors from a truncated singular value decomposition. Only the specified number of modes are computed to minimize the computational footprint. A randomized SVD can also be computed to cut costs even further.
        \item \texttt{RandomProjection}: Multiply measurements with random Gaussian vectors to project them to a new space. This basis is related to compressed sensing approaches~\cite{Candes2006cpam,Donoho2006ieeetit,Baraniuk2007ieeespm}. 
    \end{itemize}
    
    Each of these classes is housed in the \texttt{basis} submodule.

\subsection{Other features}
    Included with \texttt{PySensors} is a large suite of examples, implemented as Jupyter notebooks.
    Some of the examples are written in a tutorial format and introduce new users to the objects, methods, and syntax of the package.
    Other examples demonstrate intermediate-level concepts such as how to visualize model parameters and performance, how to combine \texttt{scikit-learn} and \texttt{PySensors} objects, selecting appropriate parameter values via cross-validation, and other best-practices.
    Further notebooks use \texttt{PySensors} to solve challenging real-world problems.
    The notebooks reproduce many of the examples from the papers upon which the package is based~\cite{manohar2018data,clark2018greedy,brunton2016sparse}.
    To help users begin applying \texttt{PySensors} to their own datasets even faster, interactive versions of every notebook are available on Binder.
    Together with comprehensive documentation, the examples will compress the learning curve of learning a new software package.

\section{Examples}\label{sec:examples}
In this section we demonstrate the use of \texttt{PySensors} classes and methods with a set of examples.
We show both reconstruction and classification problems. Additional examples are available on the \texttt{PySensors} documentation site\footnote{\url{https://python-sensors.readthedocs.io/}}.

\subsection{Reconstruction examples}
Consider the problem of interpolating a real-valued function $f$ on $[0, 1]$. Suppose we wish to use polynomials to perform this interpolation. The sparse sensor placement problem is then equivalent to the problem of selecting points in $[0,1]$ at which to sample $f$. Before we can choose these optimal interpolation points, we must choose a polynomial basis to use. Ideally we should work with a numerically stable basis such as Chebyshev polynomials, but suppose we choose something simple such as monomials:

\begin{equation}
    \Psi_r = \bbm \horzline & 1 & \horzline \\ \horzline & x & \horzline \\ \horzline & x^2 & \horzline \\ & \vdots & \\ \horzline & x^{r-1} & \horzline \ebm
\end{equation}

\begin{lstlisting}[language=iPython]
import numpy as np

x = np.linspace(0, 1, 1001)
r = 11
psi_r = np.vander(x, r, increasing=True).T
\end{lstlisting}

In keeping with \texttt{Scikit-learn} conventions, each row corresponds to an example and each column to a feature or sensor location. This choice departs from the mathematical literature wherein examples are typically represented as column vectors.
We can then use the \texttt{SSPOR} class to find close-to-optimal sensor locations tailored to this basis. The \texttt{fit} method is used to learn the locations.
\begin{lstlisting}[language=iPython]
from pysensors.reconstruction import SSPOR

selector = SSPOR()
selector.fit(phi_r)
\end{lstlisting}

The \texttt{SSPOR} object now contains a list of sensor locations ranked in descending order of importance stored in its \texttt{ranked\_sensors\_} attribute. Note that because the model was fit on 11 examples, only the first 11 sensor locations are meaningful. The remaining indices are given in random order. We can specify the number of sensors we would like to see via the \texttt{set\_n\_sensors} function:
\begin{lstlisting}[language=iPython]
selector.set_n_sensors(10)
print(x[selector.selected_sensors])
\end{lstlisting}
which prints the following locations
\begin{verbatim}
[1.    0.641 0.    0.884 0.289 0.47  0.099 0.958 0.763 0.036]
\end{verbatim}
These approximate the Fekete (or Gauss–Lobatto) points, which are known to be optimal for interpolation via monomials~\cite{fejer1932bestimmung}.
Alternatively we could have initialized the object with this preference \texttt{SSPOR(n\_sensors=10)}. 

\begin{figure}[t]
    \centering
    \includegraphics[width=\textwidth]{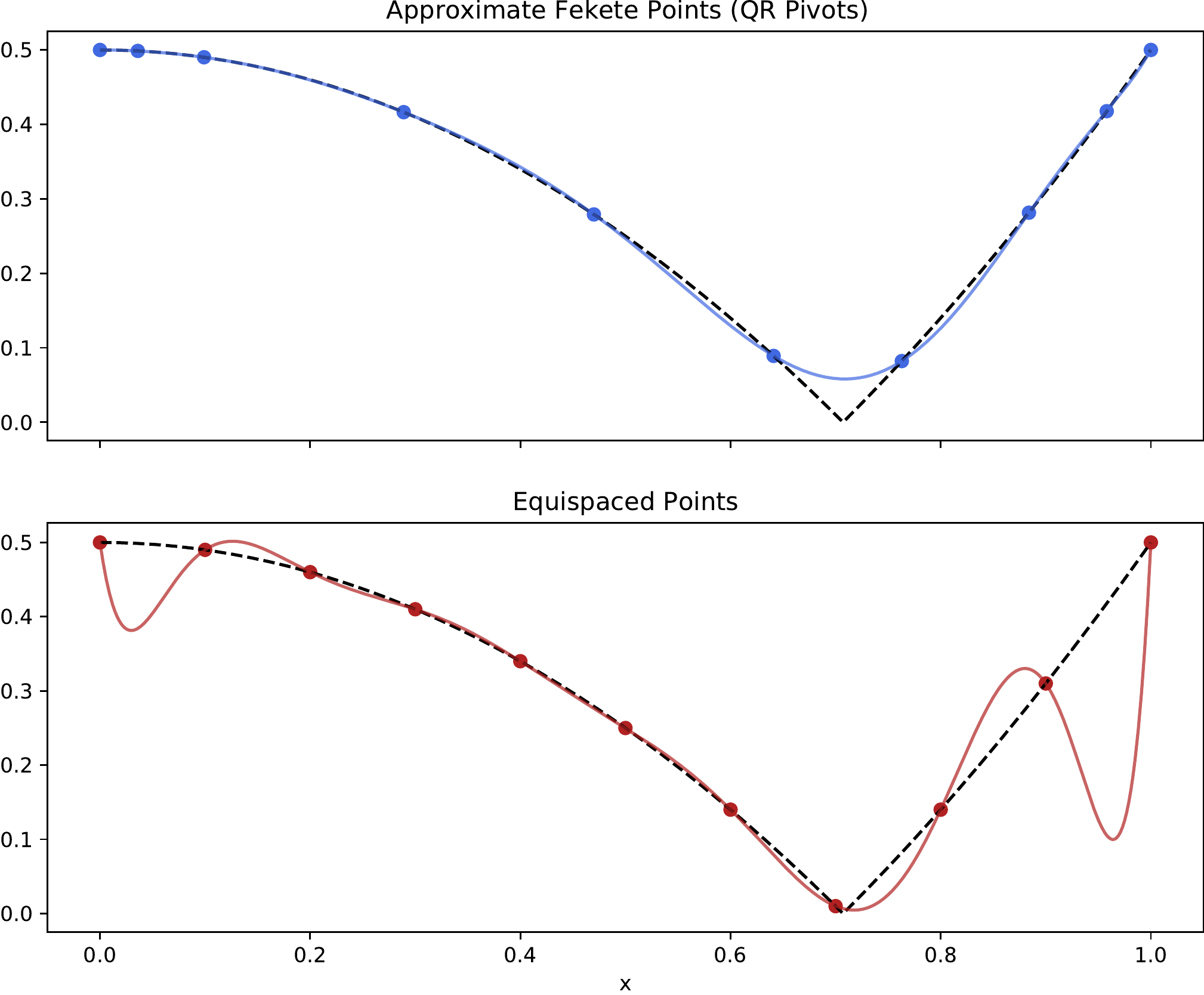}
    \caption{Reconstruction: comparison of polynomial interpolants (solid) of $f(x) = \left|x^2-\tfrac12\right|$ (dashed) obtained with  interpolation points chosen by a \texttt{SSPOR} object (top) and equispaced points (bottom).}
    \label{fig:reconstruction}
\end{figure}

The fitted \texttt{SSPOR} object can also be used to reconstruct the signal (perform interpolation) from sparse measurements. The \texttt{predict} function is used to accomplish this. We will apply the method to a function not in the range of the monomial basis (see Figure S6 of \cite{manohar2018data}):
\begin{equation}
    f(x) = \left|x^2 - \frac12\right|.
\end{equation}
\begin{lstlisting}[language=iPython]
f = np.abs(x ** 2 - 0.5)
f_interp = selector.predict(f[selector.selected_sensors])
\end{lstlisting}
For comparison, in Figure~\ref{fig:reconstruction} we plot the interpolants generated by both our method and equispaced points.
\begin{lstlisting}[language=iPython]
from numpy.linalg import lstsq

equi = np.arange(0, 1001, 100)
equi_interp = np.dot(phi_r, lstsq(phi_r[equi, :], f[equi])[0])
\end{lstlisting}

\begin{figure}[t]
    \centering
    \includegraphics[width=\textwidth]{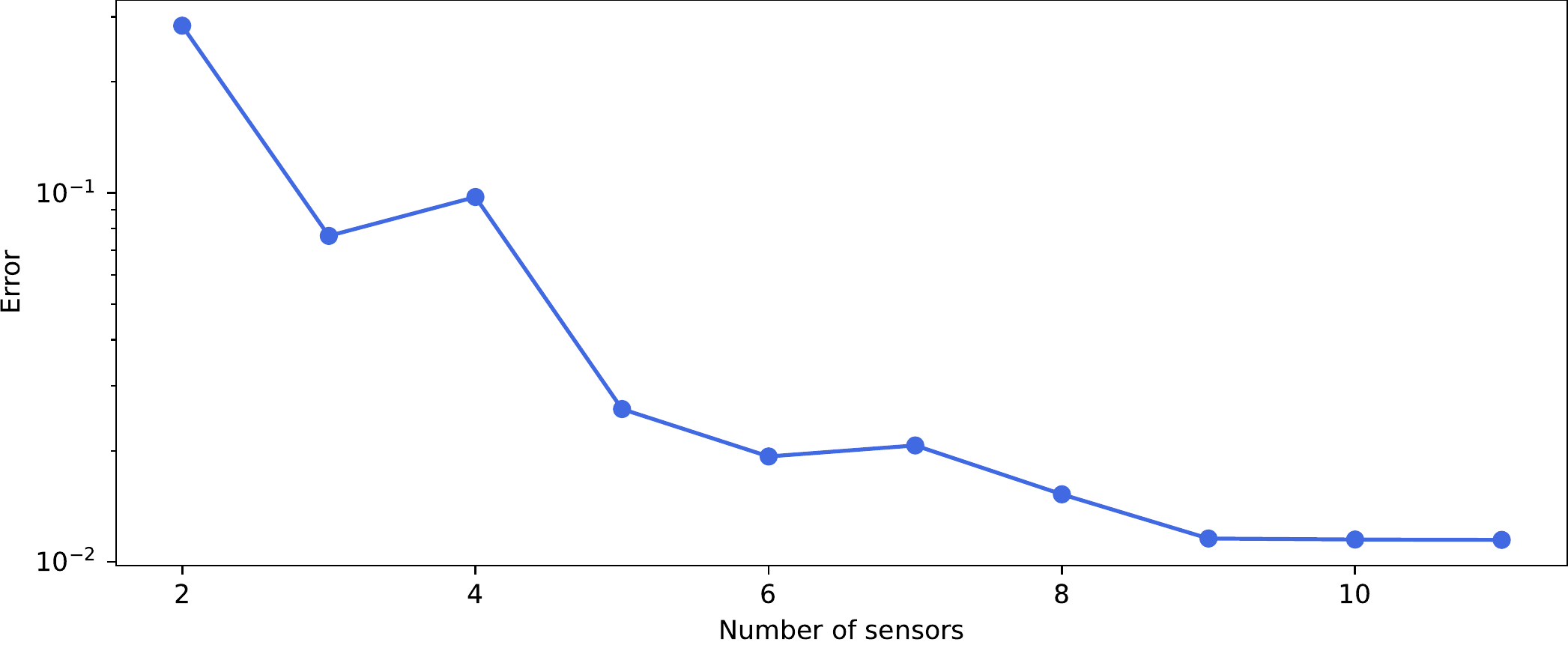}
    \caption{Reconstruction: root-mean-square error in the \texttt{PySensors} reconstruction of $f(x) = \left|x^2-\tfrac12\right|$ as a function of the number of sensors.}
    \label{fig:reconstruction-error}
\end{figure}

A common question to ask at this point is ``How does the reconstruction error depend on the number of sensors I use?'' We can use the \texttt{reconstruction\_error} method to get an answer. We simply specify an array of values of \texttt{n\_sensors} to try and the test data on which to measure the error and the function will compute a metric of interest for each number of sensors. The default metric or scoring function is the root-mean-square error.
\begin{lstlisting}[language=iPython]
sensor_range = np.arange(2, r + 1)
recon_error = selector.reconstruction_error(f, sensor_range)
\end{lstlisting}
A plot of the reconstruction error is given in Figure~\ref{fig:reconstruction-error}.

\subsection{Classification examples}
\begin{figure}[t]
    \centering
    \begin{subfigure}[t]{0.5\textwidth}
        \includegraphics[width=0.9\linewidth]{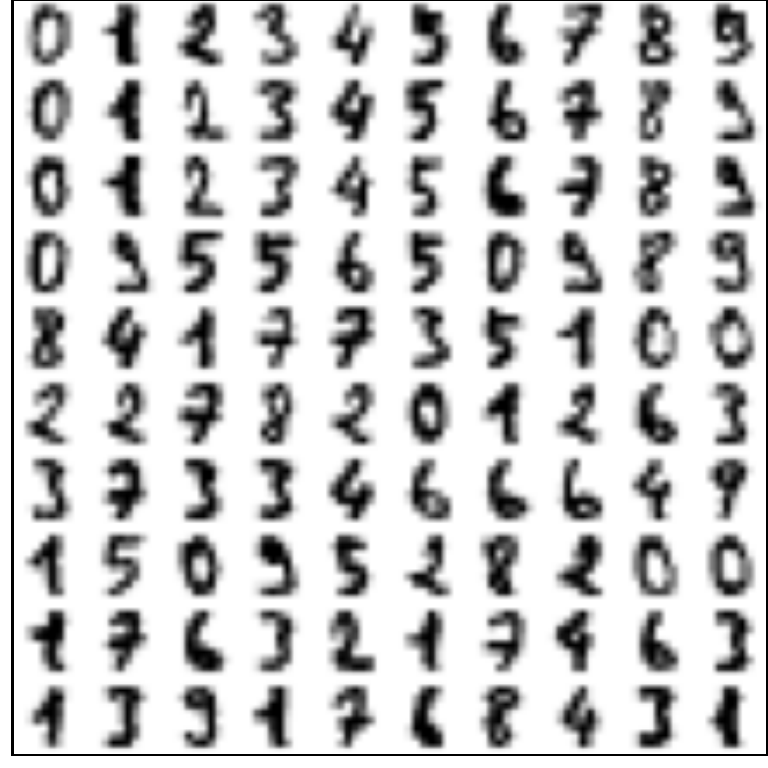}
        \caption{Classification: examples of images from the digits dataset.}
        \label{fig:digits}
    \end{subfigure}%
    \begin{subfigure}[t]{0.5\textwidth}
        \includegraphics[width=0.9\linewidth]{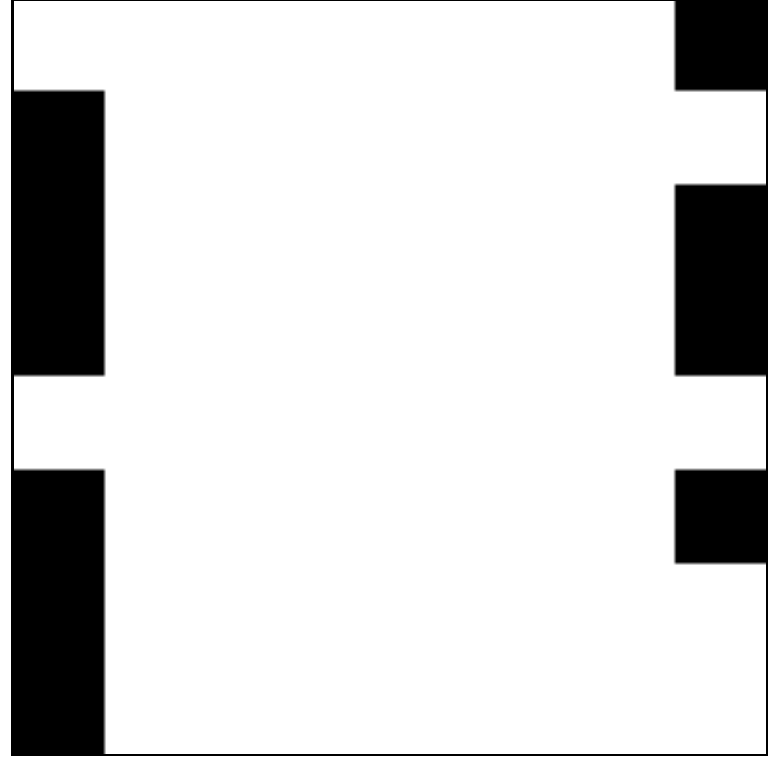}
        \caption{Pixels selected by \texttt{SSPOC} using LDA.}
        \label{fig:digits-sensors}
    \end{subfigure}
    \caption{Results for the digits dataset.}
\end{figure}

Next we turn to the problem of identifying a sparse set of sensor positions optimized for \textit{classification} tasks. For the sake of simplicity, we will work with the digits dataset of \texttt{Scikit-learn}, which consists of eight-by-eight images of handwritten digits. See Figure~\ref{fig:digits} for example images. Our overall objective is to train a classifier to predict which digit is drawn in each image. We will add the restriction that the classifier is limited in the pixels it is allowed to see.
The class designed to select the most salient sensor locations is named after the algorithm it employs: Sparse Sensor Placement Optimization for Classification, or \texttt{SSPOC} for short.
\texttt{SSPOC} instances can be used in much the same way as standard \texttt{Scikit-learn} estimators. They are fit to the data with a \texttt{fit} method:
\begin{lstlisting}[language=iPython]
from pysensors.classification import SSPOC

classifier = SSPOC(n_sensors=10)
classifier.fit(X_train, y_train)
\end{lstlisting}
and output predictions via \texttt{predict}:
\begin{lstlisting}[language=iPython]
y_pred = classifier.predict(X_test[:, classifier.selected_sensors])
\end{lstlisting}
Note that once the estimator has been fit it expects subsequent samples to consist only of measurements taken at the sensors it has chosen. This is because it refits its internal classifier---in this case linear discriminant analysis (LDA)---on the subsampled data upon deciding on a set of sensor locations. Figure \ref{fig:digits-sensors} visualizes the 10 pixels that were picked. Which pixels optimize classification accuracy depends on the classifier that is used. The \texttt{PySensors} SSPOC implementation is compatible with any linear classifier.

The number of sensors can be modified after fitting, but the training data are needed to refit the classifier for the new set of sensors:
\begin{lstlisting}[language=iPython]
classifier.update_sensors(n_sensors=5, xy=(X_train, y_train))
\end{lstlisting}

There are many other parameters affecting the performance of the \texttt{SSPOC} class that we omit from this discussion. Please see the documentation and examples for a more comprehensive exploration of such options.

\subsection{Basis examples}
Both the \texttt{SSPOR} and \texttt{SSPOC} constructors accept a \texttt{basis} parameter, which specifies the basis in which to represent the data:
\begin{lstlisting}[language=iPython]
# SVD basis for a SSPOR model
svd = pysensors.basis.SVD(n_basis_modes=10)
selector = SSPOR(basis=svd)

# Random projection basis for a SSPOC model
rp = pysensors.basis.RandomProjection(n_basis_modes=30)
classifier = SSPOC(basis=rp)
\end{lstlisting}

It is often useful to track how the performance of the model changes as the number of basis modes is varied. One option would be to refit the model for each number of basis modes. However, for bases such as \texttt{SVD}, which are computed based on input data, this approach is wasteful. Each time the fit method is called, the SVD modes will be recomputed. For this reason \texttt{PySensors} has convenience functions for efficiently updating the number of basis modes. They allow us to avoid refitting the basis, but not re-computing the sensor locations. The methods are slightly different for \texttt{SSPOR} and \texttt{SSPOC} objects in that the training data are required for \texttt{SSPOC} but not for \texttt{SSPOR}:
\begin{lstlisting}[language=iPython]
# SSPOR object
selector.fit(phi_r)
selector.update_n_basis_modes(5)

# SSPOC object
classifier.fit(X_train, y_train)
classifier.update_n_basis_modes(20, xy=(X_train, y_train)
\end{lstlisting}
It is important to start with a large number of basis modes, then update to smaller numbers of basis modes.

\section{Practical tips}\label{sec:practical-tips}

\subsection{Reconstruction}
It is important to select the appropriate basis for a given problem, such as choosing polynomials for function approximation in the example above. For a general large, data-driven problem such as sea surface temperatures or photographs of faces, the identity basis (performing QR on the raw snapshots) will produce the lowest reconstruction error at a given number of sensors. This is because no information is lost to construct a low-rank approximation of the data, but by that same virtue, the identity basis can lead to impractically long run times for a large data set. Hence the built-in options to use SVD or randomized projections, both of which provide optimal or near-optimal low-rank approximations. Other basis options that could be manually employed include the dynamic mode decomposition basis, Fourier modes, and basis modes arising from the solution of a system's equations of motion, if known.

In general, the reconstruction error will decrease as the number of sensors and basis modes increases, but the number of sensors relative to the number of modes is also important and depends on the basis. With an SVD basis, as the number of modes is increased, any additive noise in the measurements begins to dominate the reconstruction error, leading to the unintuitive result that reconstruction error increases with the number of modes, also see \cite{peherstorfer2020stability}. This can be mitigated by oversampling, i.e. using more sensors than modes. Note that when oversampling, \texttt{PySensors} randomly selects the sensors beyond the number of modes $r$, which has been shown to be fast and effective \cite{peherstorfer2020stability, clark2020multi}.

Conversely, with random projections, the quality of the reduced-order approximation depends on the presumed rank of the system \cite{liberty2007randomized,halko2011finding}. In order for the randomized approximation to have a high probability of accurately representing the system, the number of modes should be at least five or ten more than the system's presumed rank. When sparsely sampling, the rank of the system can be at most equal to the number of sensors $p$, and so we recommend choosing at least $p+10$ basis modes.

If using a cost function, determine the trade-off between cost savings and reconstruction accuracy by multiplying the cost function by a constant factor. If this factor is set to zero, unmodified QR is performed and the sensor locations with the lowest reconstruction error will be returned. If the weighting is large (what constitutes ``large" depends on the system, basis, and cost function), low-cost sensors will be selected, regardless of their effectiveness for reconstruction.

\subsection{Classification}

Just as for reconstruction, the choice of an appropriate low-dimensional basis for the data of interest is crucial.
Generally, the SVD or a random projection produce low-rank approximations that are nearly optimal for most datasets.
The choice of $r$, the rank of the projection, determines the number of sensors chosen.
For binary classification, the number of sensors chosen is approximate $r$; for $c>2$, the number of sensors is at most $r(c-1)$.

For certain datasets, some reweighing of the bases improves the performance of the sparse classification.
When the most discriminating directions in the data are not well captured by the most energetic modes, it is helpful to construct a biased basis by reweighing the basis vectors by largest magnitude elements in $\mathbf{w}$~\cite{Mohren2018pnas}.
In other words, instead of using the largest singular values $\mathbf{\Sigma}_r$ to choose the $\mathbf{\Psi}_r$ basis, we may start with a larger $r$, compute $\mathbf{w}$, and then use the largest elements of $\mathbf{\Sigma}_r |\mathbf{w}|$ to learn a modified SVD basis better tailored to the training data.

Since the learning of sparse sensor locations and the execution of classification on those sparse sensors are separate, decoupled tasks, we typically re-train a separate classifier on the sparse sensor locations.
While the theoretical derivation of SSPOC relies on linear projections and discriminates, this ultimate step can take many forms, including any nonlinear classifier.

\section{Acknowledgments}
The authors acknowledge support from the Air Force Office of Scientific Research (AFOSR FA9550-19-1-0386) and The Boeing Corporation.  

\newpage
\begin{spacing}{.9}
  \small{
    \setlength{\bibsep}{6.5pt}
\bibliographystyle{IEEEtran}
    \bibliography{arxiv}
  }
\end{spacing}

\end{document}